# PROBLEM REDUCTION IN ONLINE PAYMENT SYSTEM USING HYBRID MODEL

Sandeep Pratap Singh[1], Shiv Shankar P. Shukla[1], Nitin Rakesh[1] and Vipin Tyagi[2]

[1]Department of Computer Science and Engineering, Jaypee University of Information Technology, Waknaghat, Dist. Solan, India
sandeep102209@gmail.com, shivshu@gmail.com, nitin.rakesh@gmail.com
[2]Department of Computer Science and Engineering, Jaypee University of Engineering and Technology, Guna, M.P., India
dr.vipin.tyagi@gmail.com

## ABSTRACT

*Online auction, shopping, electronic billing etc. all such types of application involves problems of fraudulent transactions. Online fraud occurrence and its detection is one of the challenging fields for web development and online phantom transaction. As no-secure specification of online frauds is in research database, so the techniques to evaluate and stop them are also in study. We are providing an approach with Hidden Markov Model (HMM) and mobile implicit authentication to find whether the user interacting online is a fraud or not. We propose a model based on these approaches to counter the occurred fraud and prevent the loss of the customer. Our technique is more parameterized than traditional approaches and so, chances of detecting legitimate user as a fraud will reduce.*

## 1. INTRODUCTION

Online shopping across the world is replacing crowded stores for its one click convenience .It has become a popular option for the consumers. According to the trends in online shopping a global consumer report February 2008, the world' s online population that use the internet for shopping is over 85%, increased by 40% from last two years and in this maximum users are regular online shoppers i.e. doing online shopping once per month . The escalating popularity of the online shopping is really a global phenomenon .South Korea with the 99% of internet users shopped online has the largest number of online shoppers in the world. German, U.K and Japanese consumer comes after that [1].

As the credit card users around the globe is going up continuously it is also providing more opportunities to the illegitimate user to commit fraud i.e. to steal the card details and subsequently making illegal transactions. The total credit card fraud in United Kingdom alone was around 535 million pound in 2006 [2] and 750 to 830 million dollar in U.S. in 2006 [3].

Fraud is one of the very serious issues in the credit card industry which may cause very severe financial loss to the consumer. So the main aim is to detect the fraud and to protect the consumers from being cheated. The fraud can be done either by stealing the physical card or get the important card details like card numbers, security code, validity etc. This type of details can be used to do a fraud on the internet. This type of fraud is not easy to detect since the genuine user may not know that his card information is used by the some other person. The most common way to detect this type of fraud is examining the patterns of the card activities and other things like mobile phone activities. Based on these patterns profile of the legitimate user will be created. As human beings mostly follow the specific behaviour pattern so any unusual deviation from the behaviour is a threat.

This paper is divided in five sections. Section first is introduction where the problem is defined and the recent trends are introduced. Second section consists of previous work in this domain. In third and fourth section the model is proposed and the case studies on this model are discussed. Section fifth is concluding this paper and is followed by references.

## 2. RELATED WORK

In the field of credit card fraud detection there is lot of research area with no of technique used especially neural network and the data mining. Dorronsoro et al (1997) [4] proposed an online fraud detection system based on neural classifier. Here the main problem was that data is needed to be grouped by the kind of the account. Card watch by Aleskerov et al [5], it is a data mining system based on neural learning module used for fraud detection. But this was inefficient due to mix detection of fraudulent transaction. To reduce the number of misdetection Kim & Kim proposed a concept [6]. Brause et al used data mining tools like Clementine and neural network technologies to obtain better fraud detection [7].

Bayesian networks can also be applied to detect fraud in credit card as proposed by Maeset. al. [8].The main disadvantage of this technique is the time limit as compared with neural network. Chiu and Tsai [9] proposed combination of the data mining techniques for banking industry, by using this bank has knowledge of the fraud patterns in distributed environment.

In most of these above techniques they need examples fraud cases due to card stealing, loss card, counterfeit fraud based on which fraudulent transactions are identified. In a real scenario to get the fraud data is a big problem and since day by day new types of fraud are appearing. To solve this problem of fixed fraud data we presents a new model which does not required fixed fraud data but can detect the fraud by cardholder spending behaviour and mobile activities. This model uses combination of two techniques one is based on spending profile i.e. HMM model and other is based on the mobile implicit authentication system. The details of this model are in section 3 and 4.

## 3. PROPOSED WORK

By this model we are combining two techniques namely HMM techniques on spending profile and mobile implicit authentication system.

### 3.1. Mobile authentication architecture

According to the proposed work by Richard Chow et al. [10] a fraud can be detected based on architecture for supporting authentication and authentication based on some behaviour pattern of the user called implicit authentication. Here this behavioural pattern is based on mobile phone data like call history, SMS activity, location, internet access etc. This behavioural pattern will be used to generate an authentication score based on past behaviour and recent behaviour and this score is compared with a threshold which decide whether to accept or reject the user. The main problem with is approach is when the mobile is stolen, in this false positive will be high (this means legitimate user is not authenticate and will be treated as fraud).

### 3.2. HMM model

According to the proposed work of Abhinav Srivastava et al. [11], hidden markov model can be used to detect fraud. HMM observes the sequence of the money spend in each transaction. Based on the spending habits of the user HMM develop spending profile of a card holder generally it uses three profile low, middle and high. After computation user is given a spending profile which suited his spending habits. Here a sequence of a card transaction is formed and after new transaction a deviation is checked. If there is deviation then percentage of deviation is compared with threshold to decide fraud. If deviation is greater than threshold then fraud else the new transaction is added to the sequence. In this approach accuracy of the system is close to 80%. Here there is a scope of improving the false positive by providing more restriction.

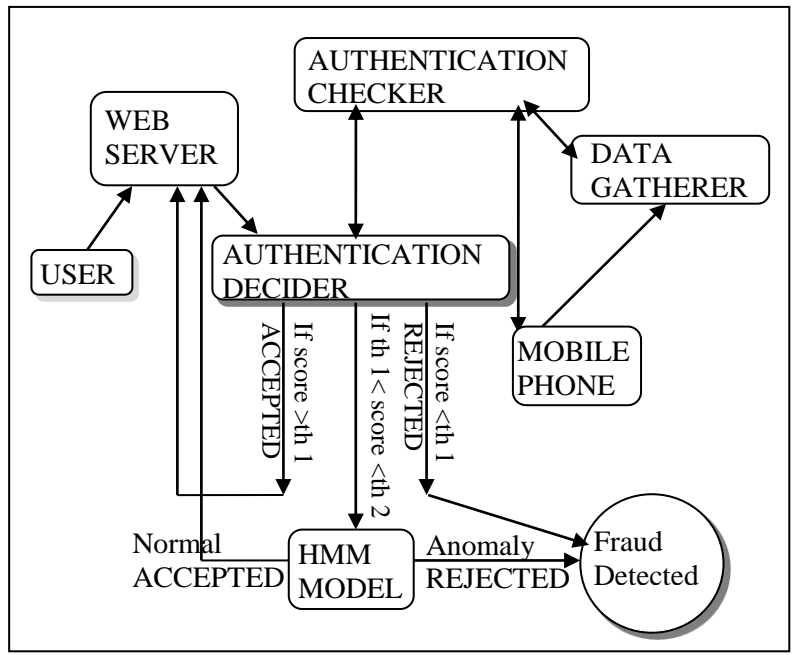

Figure 1.Proposed Hybrid Model

### 3.3. Hybrid model

By this proposed model, see figure 1, we are combining the two approaches as discussed above i.e. HMM technique and the authentication by the user mobile.

Table 1. Algorithm for Proposed Hybrid Model.

| Step 1 | User accesses the web server and makes the payment request to the authentication decider. |
|---|---|
| Step 2 | Authentication decider asks for the authentication details from the authentication checker. |
| Step 3 | Authentication checker refer to mobile device and /or data gatherer for the authentication details based on mobile behavioural pattern. |
| Step 4 | Authentication checker then generate the score (discussed below) and send it to authentication decider. |
| Step 5 | Authentication decider then compare score with the thresholds than follow any of the three cases discussed below. |

The detail explanation of the complete process is discussed in this section and with the help of the case studies in the next section.

In this architecture we consider the following types of blocks i.e. mobile phone, data gatherer, authentication checker and authentication decider, HMM model block(this block includes complete HMM model which is based on spending habits of consumer as discussed in part B, [11]). In this approach user made a payment request (for any article, commodity) from a web page to the authentication decider. The data gatherer collects the mobile data from mobile information like SMS history, phone call history, history of browser, information of the network and device location regularly.

The authentication checker takes the data from the data gatherer or directly from the client device. The authentication checker makes all the decision with the help of the data collected and the policies of authentication. These policies are given to the authentication by the authentication decider which is obtained from the user request (payment request). After that authentication decider compare the probability score obtained from authentication checker with the two threshold namely threshold 1 and threshold 2 (assuming: threshold 1 < threshold 2).

The score mention above is the probability of matching the behavior, it is computed with the past and recent behavior of the user .the thresholds can differ for different applications and there security parameter. The comparison of score and threshold is used for making decision for authentication. If the score is less than threshold 1 then request will be directly rejected. This means there is high degree of mismatching in the pattern of behaviour so there is definite chance of a fraud. The figure 2 shows how model will flow in this case.

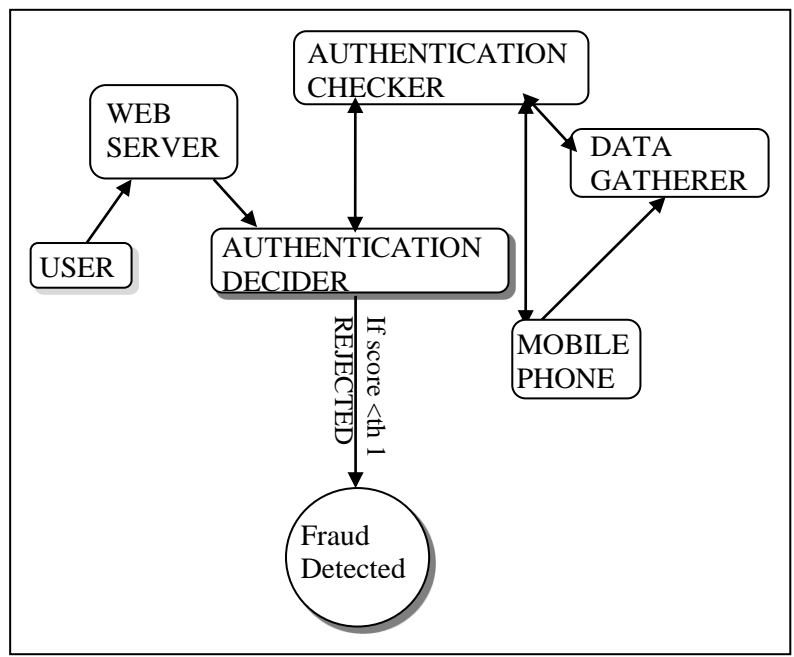

Figure 2. Flow diagram for case 1.

If threshold 1 < score < threshold 2 in this case there is slight or partial matching of the pattern so the requested transaction is forwarded to the hmm model .now if according to the HMM model there is deviation or anomaly in the spending profile of the card holder then alarm is raised else if everything is normal then request is fulfilled. The figure 3 shows how this model flows in this case. If score >threshold 2 then the request is directly accepted because there is high degree of matching probability i.e. pattern behavior is almost same as the legitimate user. The figure 4 shows how this model flows in this case. The mobile phone operating system collects the data that includes user's activities on the mobile like phone call patterns, SMS activities, location, web access etc.

After this it informs this to data gatherer on a regular basis. These data are temporally stored till data gatherer obtains it completely and then it is erased from the local memory of the mobile. This exchange of data is done securely with the help of some encryption techniques applied before the exchange of data. This is done to protect the privacy of the customer. In this way the patterns will be detected while keeping in the mind the privacy of the customer.

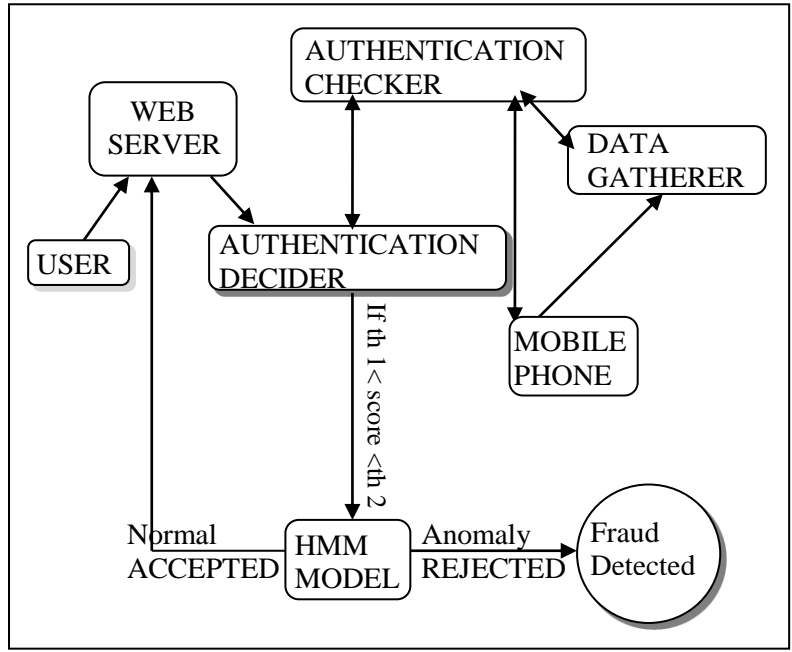

Figure 3. Flow diagram for case 2.

The authentication checker collects the data from the data gatherer and sometimes directly from mobile device. It collects the data based on the policy provided by the authentication decider. When a payment request comes to authentication decider it forward the request to the authentication checker and the details associated with it authentication checker then send the query to the mobile device and/or data gatherer.

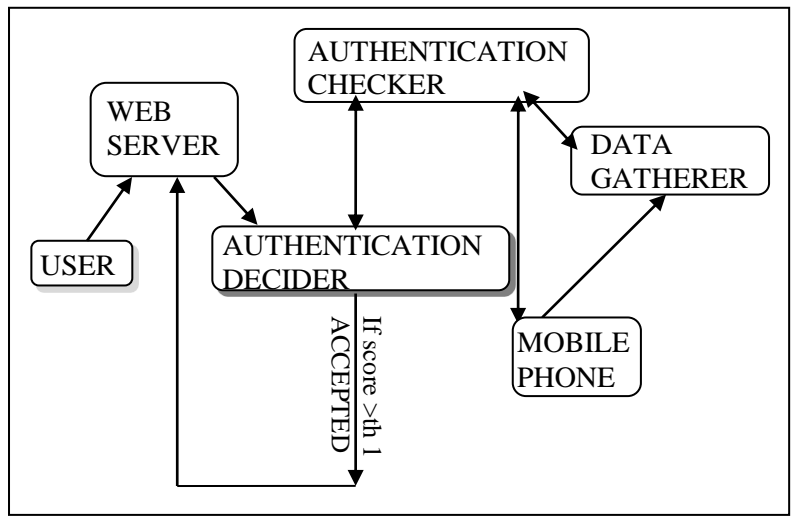

Figure 4. Flow diagram for case 3.

The data gatherer respond to the query and send back the required details to authentication checker. It then computes the authentication result based on the policy .the authentication checker also calculate the score of the matching pattern with help of the past and recent behavior. This score is send to authentication decider where it compares it with the threshold 1 and threshold 2 as mentioned above.

## 4. CASE STUDY AND RESULTS

In our proposed model here we study it on two scenarios namely mobile theft and card theft (details).

### 4.1. Mobile theft

By mobile authentication approach we have to only relied on the pattern observe by the authentication system. So in this case there is high probability that the legitimate user will be treated as fraudulent user. HMM model is not affected by mobile theft. So mobile theft has no significance here it only depends on the observation of the card holders spending behaviour. In Hybrid model we do not only rely on the pattern of the mobile details. Here we use both the pattern of the mobile and the spending behaviour of the card holder depending on the probability of the pattern matching by the first phase. The possible cases are as follows:

**Case 1**: High profile: this means the pattern observe in the first phase has a high probability of matching .i.e. authentication score > threshold 2. But in case of mobile theft there is minimal chance of high score because the thief usage patterns will definitely differ from legitimate user pattern.

**Case 2**: Middle profile: this means the pattern observe in the first phase has a probability of partial matching .i.e. threshold 1 < authentication score < threshold 2. In case of mobile theft it is the most probable case because only mobile patterns can't give the surety of mobile theft so our model will further check the transactions with help of $2^{nd}$ phase i.e. HMM phase. In this way we are more probable than the previous method i.e. mobile authentication approach, to detect the fraud user and chance of the false positive (to detect the legitimate user as fraud user) is less.

**Case 3**: Low profile: this means the pattern observe in the first phase has a very low probability of matching i.e., authentication score < threshold 1. In case of the mobile theft this also a probable case but for this the thief observation patterns should be more so that there is high mismatching of the patterns. In this case the user will be identity as a fraud and can be directly rejected. Example is that the mobile of user has been stolen. Now, let us suppose our threshold 1 is at 25% and threshold 2 is at 75%.

In this case since mobile is stolen so call patterns, location and others things pattern will deviate from actual therefore pattern mismatching will b there and there is very little chance that authentication score i.e. percentage of matching will be higher than threshold 2 i.e. 75%.the maximum a probability is of the middle case in which score is between threshold 1 and threshold 2 i.e. 25 and 75.here it will again the user for its authenticity through hmm model, which depends on spending profile.

And if percentage of matching is very less then will directly reject the user i.e. authentication score or percentage of matching is very less let's say below 25. In all there is very less change that we treating legitimate user as fraud user therefore our checking is much stricter than other approaches. For example let us take records of a mobile use for 1 week. Here we consider only 2 parameters; they are no. of phone calls and number of SMS see table 2.

Table 2. Records of calls and SMS.

|  | Sun | Mon | Tue | Wed | Thurs | Fri | Sat |
|---|---|---|---|---|---|---|---|
| *No. of calls* | 5 | 10 | 15 | 3 | 3 | 4 | 5 |
| *No. of SMS* | 10 | 15 | 4 | 2 | 3 | 2 | 3 |

Now let's on next day mobile is being lost or stolen, we are considering if mobile is stolen then number of calls and number of SMS will be zero. Let us suppose that threshold 1 is 25 and threshold 2 is 75. Now we will calculate the score based on these patterns at a time we are considering 7 days. So initial average no. call is as follows

$$\text{Avg initial} = \frac{(x1 + x2 + x3 + x4 + x5 + x6 + x7)}{7}$$

$$= \frac{(5 + 10 + 15 + 3 + 3 + 4 + 5)}{7}$$
$$= 6.42$$

Now on next day number of call is zero so new average will be calculated based on last 7 days.

$$\text{New avg} = \frac{(10 + 15 + 3 + 3 + 4 + 5 + 0)}{7}$$
$$= 5.71$$

Score = percentage of matching = [100 − ((previous average − recent average) / previous average) × 100]

$$= \left[100 - \left(\frac{(6.42 - 5.71)}{6.42}\right) \times 100\right]$$
$$= 88.94$$

For number of SMS we can calculate the score as follows:

$$\text{Avg initial} = \frac{(x1 + x2 + x3 + x4 + x5 + x6 + x7)}{7}$$

$$= \frac{(10 + 15 + 4 + 2 + 3 + 2 + 3)}{7}$$
$$= 5.57$$

Now on next day number of call is zero so new average will be calculated based on last 7 days.

$$\text{New avg} = \frac{(15 + 4 + 2 + 3 + 2 + 3 + 0)}{7}$$
$$= 4.14$$

Score = percentage of matching = [100 − ((previous average − recent average) / previous average) × 100]

$$= \left[100 - \left(\frac{(5.57 - 4.14)}{5.57}\right) \times 100\right]$$
$$= 74.32$$

Now the score of parameters are averaged, i.e.

$$= \frac{(88.94 + 74.32)}{2} = 82.13$$

As per our model score is compared with thresholds. Here score is 82.13 which is greater than threshold 2 i.e. 75. So this will be accepted as per proposed hybrid model. Now on next to next day number of the call and the number of SMS will be again zero. Thus new average for number of calls = 4.28 and the previous average = 5.71. Therefore by using above mentioned formula score = 75. The new average for the number of the SMS = 2 and previous average = 4.14. Therefore score = 49.4. Thus overall average score = 62.2. This score is between threshold 1 and threshold 2, so user will be further authenticated by HMM model. This score will further degrade as the number of the day's increases and after the score is below threshold 1 the request will be directly rejected. The above discussed example can be further improved by considering more parameters

### 4.2. Credit Card Stolen (card detail)

By mobile authentication the location of the card transaction and the mobile device is more probable to differ. So there is a probability of fraud. By hmm model: in this case if the card details are stolen the spending behaviour of the user will deviate from the actual behaviour so there is the probability of the fraud. In this case we are only looking the spending behaviour which may not be enough to detect the fraud. By the hybrid model: In our model we use both the pattern of the mobile and the spending behaviour of the card holder depending on the probability of the pattern matching by the first phase. There are three possible cases:

**Case 1**: High profile: this means the pattern observe in the first phase has a high probability of matching .i.e. authentication score > threshold 2. But in the case of card stolen the location of the card and the mobile may differ so there is very less chance of pattern matching .so this case may not be applied in card stolen.

**Case 2**: Middle profile: this means the pattern observe in the first phase has a probability of partial matching .i.e. threshold1 < authentication score < threshold 2. In case of card stolen at the initial period the pattern mat5ching will be partial so we can't say directly whether it is fraud or legitimate user. So the detection will be further carried out by the $2^{nd}$ phase i.e. hmm model, which deals with spending behaviour of the user. So there is better chance of detecting the fraud.

**Case 3**: Low profile: this means the pattern observe in the first phase has a very low probability of matching .i.e. authentication score < threshold 1. In case of card stolen after a certain period there will be high mismatching in the pattern where location (i.e. location of the card transaction and the mobile device) will be the prime factor. So we need not to move to the $2^{nd}$ phase and can be directly reject the user treating him as a fraud.

### 5. CONCLUSION

In this paper we have proposed a model which combination of hmm model based on spending profile of user and mobile implicit authentication based on mobile data patterns. this model impose much security to the cardholder as it has taken two things under consideration (spending habits and mobile usage habits).in this is less chances that legitimate user will be treated as fraud, means false positive will be decreased. We will extend this approach for different other online frauds and their explicit detection techniques. It is still required to specify the technique of fraud detection with respect to a fraud. This type of detection model which combines complete detection and prevention mechanism will be more productive to online communication and commerce.

**Sandeep Pratap Singh** received the B.E. degree in Information Technolgy from Lakshmi Narain College of Technology, Bhopal in 2009.He is currently working towards the MTech degree in Jaypee University of Information Technology, Waknaghat, Solan-173215.His research interest include fraud detection, communication and network security and coding theory.

**Shiv Shankar Prasad Shukla** received the Diploma's degree in computer application from Government Polytechnic Mining Institute, Dhanbad (Jharkhand) in 2007. He holds Bachelor's Degree in computer technology from Priyadharshini College of Engineering, Nagpur (Maharastra) in 2010. He is currently working towards the MTech degree in Jaypee University of Information Technology, Waknaghat, Solan-173215. His research interest include fraud detection, Cloud computing, computer and network security and information theory.

**Nitin Rakesh** is Sr. Lecturer in the Depart. of CSE, JUIT, Waknaghat, Solan–173215, India. In 2004, he received the Bachelor's Degree in Information Technology and Master's Degree in Computer Science and Engineering in year 2007. Currently he is pursuing his doctorate in Computer Science and Engineering and his topic of research is parallel and distributed systems. He is a member of IEEE, IAENG and is actively involved in research publication. His research interest includes Interconnection Networks & Architecture, Fault–tolerance & Reliability, Networks–on–Chip, Systems–on–Chip, and Networks–in–Packages, Network Algorithmics, Parallel Algorithms, Fraud Detection. Currently he is working on Efficient Parallel Algorithms for advanced parallel architectures.

**Dr. Vipin Tyagi** is Associate. Prof. in Computer Science and Engineering at Jaypee University of Engineering and Technology, Guna, India. He holds Ph.D. degree in Computer Science, M.Tech. in Computer Science and Engineering and MSc in Mathematics. He has about 20 years of teaching and research experience. He is an active member of Indian Science Congress Association and President of Engineering Sciences Section of the Association. Dr. Tyagi is a Life Fellow of the Institution of Electronics and Telecommunication Engineers. He is a senior life member of Computer Society of India. He is member of Academic-Research and Consultancy committee of Computer Society of India. He is elected as Executive Committee member of Bhopal Chapter of Computer Society of India and M.P. and CG chapter of IETE. He is a Fellow of Institution of Electronics and Telecommunication Engineers, life member of CSI, Indian Remote Sensing Society, CSTA, ISCA and IEEE, International Association of Engineers. He has published more than 50 papers in various journals, advanced research series and has attended several national and international conferences in India and abroad. He is Principal Investigator of research projects funded by DRDO, MP Council of Science and Technology and CSI. He has attended many advanced courses in various engineering discipline. He has been a member of Board of Studies, Examiner Member of senate of many Universities. His research interests include Image Processing, Pattern Recognition and Digital Forensics.